\documentclass[prb,amsmath,amssymb,twocolumn]{revtex4-1}
\usepackage{graphicx}
\usepackage{epsfig}
\usepackage{dcolumn}
\usepackage{bm}
\usepackage{rotating}
\usepackage[table]{xcolor}
\usepackage[english]{babel}
\newcommand{\colora}[1]{{\color{black} #1 }}

\usepackage[normalem]{ulem}

\begin{document} 

\title{Theory of electron-plasmon coupling in semiconductors}
 
\author{Fabio Caruso} 
\author{Feliciano Giustino} 
 
\affiliation{Department of Materials, University of Oxford, Parks Road, Oxford, OX1 3PH}
\date{\today}

\date{\today}
\begin{abstract}
The ability to manipulate plasmons is driving new developments
in electronics, optics, sensing, energy, and medicine. Despite the massive momentum of
experimental research in this direction, a predictive quantum-mechanical framework for
describing electron-plasmon interactions in real materials is still missing. Here,
starting from a many-body Green's function approach, we develop an \textit{ab initio} 
approach for investigating electron-plasmon coupling in solids.
As a first demonstration of this methodology, we show that electron-plasmon scattering
is the primary mechanism for the cooling of hot carriers in doped silicon,
it is key to explain measured electron mobilities at high doping, and it
leads to a quantum zero-point renormalization of the band gap in agreement with experiment. 
\end{abstract}
\maketitle

Plasmons are collective oscillations of electrons in solids that can exist even in the absence of 
an external driving field. During the last decade plasmons generated tremendous interest 
owing to the rise of plasmonics, the science of manipulating light and light-matter
interactions using surface plasmon polaritons \cite{lal2007}.
Plasmonic materials and devices show exceptional promise in the areas of nanoelectronics \cite{walters/2010},
photovoltaics \cite{Atwater,brongersma2015}, and radiation treatment 
therapy \cite{Lukianova2013,zhang2013}. While 
the electrodynamic laws governing plasmonics at macroscopic length-scales are well 
understood \cite{maier2007plasmonics,pitarke2007}, 
little is known about the interaction of plasmons with 
matter at the atomic scale. For example questions pertaining the interaction between 
plasmons and charge carriers in semiconductors have not been addressed on quantitative
grounds, yet they are critical to engineering materials for semiconductor plasmonics \cite{ozbay2006}.
Up to now microscopic quantum-mechanical theories of electron-plasmon interactions have 
been limited to idealised models of solids, such as the homogeneous electron gas \cite{Bohm1953,pines1962}.
While these models laid the theoretical foundations of the theory, they are not suitable for 
predictive calculations. 

In this work we introduce a first-principles method to study electron-plasmon 
coupling in solids. As a first application we focus on doped semiconductors, where the manifestations 
of electron-plasmon coupling are most spectacular. In contrast to metals and insulators, doped 
semiconductors can sustain `thermal plasmons', that is plasmons with energies comparable 
to those of lattice vibrations. 
Under these conditions electron-plasmon interactions can modify carrier lifetimes, mobilities,
and optical gaps in a manner similar to electron-phonon interactions.
Using this method we find that, in the case of degenerate $n$-type 
silicon, thermal plasmons lead to ultrafast relaxation of hot carriers, provide the main 
bottleneck to carrier mobility, and induce a zero-point renormalization of the band gap that 
exceeds the phonon-induced renormalization.

  \begin{figure*}
  \begin{center}
  \includegraphics[width=0.9\textwidth]{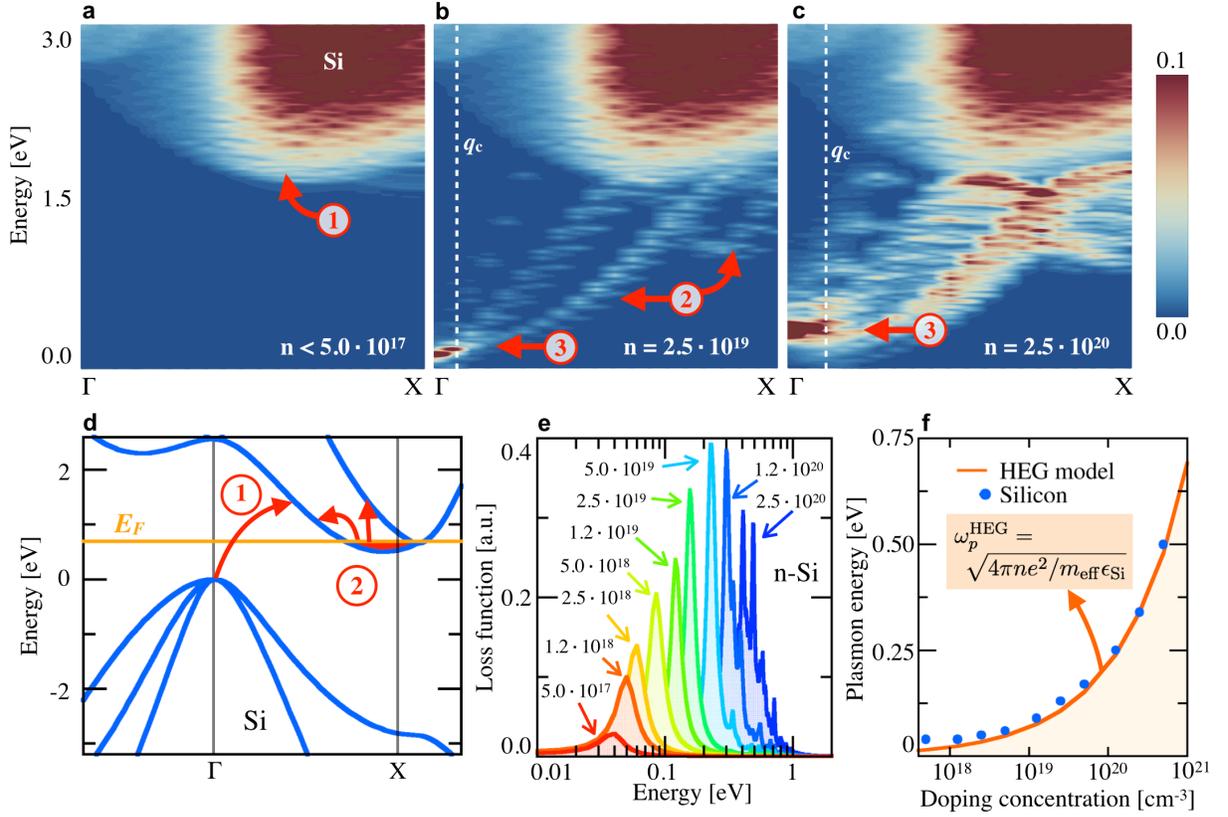}
  \caption{\label{fig1}
  (a--c) Calculated electron-energy loss function of $n$-type silicon for momentum transfers
  {\bf q} along the $\Gamma X$ high-symmetry line. The carrier density increases from left to right,
  from 10$^{17}$ to 10$^{20}$~cm$^{-3}$. 
  { (d) LDA band structure of silicon, and Fermi level ($E_{\rm F}$) for $n=2.5\cdot10^{20}$~cm$^{-3}$. }
  The step-like structures in (b) and (c)
  are only a numerical artifact arising from the limited Brillouin-zone sampling. 
  (e) Variation of the plasmon peak in the loss function vs. carrier density, evaluated at ${\bf q}=0$.
  (f) Plasma energies extracted from peaks in (e), plotted vs. carrier concentration (blue dots).
  The red line corresponds to the analytical result obtained for a homogeneous electron gas with the
  calculated isotropic effective mass and dielectric constant of silicon
  ($m_{\rm eff}=0.25$, $\epsilon_{\rm Si}=12$).}
  \end{center}
  \end{figure*}

In free-electron metals the energy of a plasmon is $\hbar\omega_{\rm P} =
(\hbar^2 e^2 n / \varepsilon_0 m_{\rm e})^\frac{1}{2}$, where $\hbar$ is the Planck constant, 
$\varepsilon_0$ is the dielectric permittivity of vacuum, and $e$, $m_{\rm e}$, and $n$ 
are the electron charge, mass, and density, respectively. At typical metallic densities, 
as in common plasmonic metals such as Au and Ag 
($n =3$-$8\cdot10^{22}$~cm$^{-3}$), 
plasmons have characteristic energies in the range of 5-10~eV. In these cases
electron-plasmon scattering is suppressed by the energy-conservation selection rule. 
At variance with this scenario, in doped semiconductors the electron mass in the above
expression is replaced by the band effective mass, and the vacuum permittivity is replaced 
by the dielectric constant. As a result the plasmon energy is considerably smaller, and 
at standard doping levels it can easily reach the thermal range, $\hbar\omega_{\rm P}=$10-100~meV.
Under these conditions electrons can exchange energy with plasmons, hence the populations 
of carriers and plasmons become mutually coupled. 

In order to investigate the consequences of this coupling, we start by characterizing plasmonic 
excitations in doped silicon from first principles. Figure~\ref{fig1} shows the calculated 
electron energy loss function, ${\rm Im}\,\epsilon^{-1}({\bf q},\omega)$, which encodes information 
about how an electron travelling through a solid dissipates energy \cite{Nozieres1959}.
Here $\epsilon^{-1}({\bf q},\omega)$ denotes the head of the inverse dielectric matrix for 
the wavevector ${\bf q}$ and the frequency $\omega$, evaluated within the random phase 
approximation \cite{adler/1962,wiser/1963}. In the case of intrinsic silicon at zero temperature 
(Fig.~\ref{fig1}a) the loss function exhibits a continuous energy distribution (brown region) 
with a threshold set by the fundamental gap. 
This broad structure arises from
interband transitions from the filled valence bands to the empty conduction bands, and physically
corresponds to the generation of electron-hole pairs by impact ionization. This is schematically 
indicated as `process 1' in Fig.~\ref{fig1}d. The scenario changes drastically in the case 
of doped silicon. Fig.~\ref{fig1}b and Fig.~\ref{fig1}c show the loss function of heavily 
$n$-doped silicon, corresponding to $n=2.5\cdot10^{19}$~cm$^{-3}$ and $n=2.5\cdot10^{20}$~cm$^{-3}$, 
respectively. As a result of the partial filling of the conduction band valley near the $X$ point 
of the Brillouin zone, new dissipation channels become available. In particular, `process 2' in Fig.~\ref{fig1}b 
corresponds to the generation of low-energy electron-hole pairs. In this case we see sharp structures 
which define `ghost' bands as a function of the momentum 
loss $\hbar{\bf q}$. These features are understood in terms of intraband and interband transitions 
from occupied initial states with wavevector ${\bf k}$ near the bottom of the conduction band to 
empty final states of wavevector ${\bf k}+{\bf q}$. The intensity of these features increases with 
the doping level from Fig.~\ref{fig1}b to Fig.~\ref{fig1}c.
The peaks in the loss function denoted by `process 3' cannot be
explained in terms of the previous two mechanisms. In fact for ${\bf q} = 0$ these structures
are much sharper than those described above, and exist below the energy (momentum) threshold
for the generation of electron-hole pairs via interband (intraband) transitions.
These processes correspond to the emission of plasmons, and are characterised by well-defined 
energy resonances, as it is shown by Fig.~\ref{fig1}e for ${\bf q}=0$. By mapping these plasmon 
peaks in the loss function we can see in Fig.~\ref{fig1}f that the plasmon energy $\hbar\omega_{\rm P}$ 
scales with the carrier concentration, following the same trend expected for a homogeneous electron gas. 
In this figure we also see that the plasmon energy is highly tunable via doping, from thermal energies at
carrier densities around 10$^{18}$~cm$^{-3}$, to half an electronvolt at densities near 10$^{21}$~cm$^{-3}$. 

At large momentum transfer $\hbar{\bf q}$ the distinction between plasmons and electron-hole
pairs is no longer meaningful, since the fluctuations of the charge density happen on length-scales
approaching the size of the crystal unit cell. In the following we identify plasmons in the loss function 
by analogy with the homogeneous electron gas, where well-defined plasma excitations exist only for
momenta below the electron-hole continuum \cite{pines1999elementary}. For a plasmon of energy 
$\hbar\omega_{\rm P}$ the critical momentum is given by the wavevector $q_{\rm c} = k_{\rm F}\left[
(1+\hbar\omega_{\rm P}/\varepsilon_{\rm F})^{1/2}-1\right]$, 
with $k_{\rm F}$ and $\varepsilon_{\rm F}$ being 
the Fermi wavevector and the Fermi energy, respectively. 
\colora{
The critical wavevector $q_{\rm c}$ marks the onset of Landau damping, that is, the decay of a plasmon upon excitation
of an electron-hole pair. Thus, for $q<q_{\rm c}$ thermal
plasmons are undamped collective phenomena with lifetimes
set by plasmon-phonon and plasmon-plasmon scattering processes \cite{sup}.
}
This boundary is shown as white dashed lines in Fig.~\ref{fig1}b and 
Fig.~\ref{fig1}c. 

  \begin{figure*}
  \begin{center}
  \includegraphics[width=0.9\textwidth]{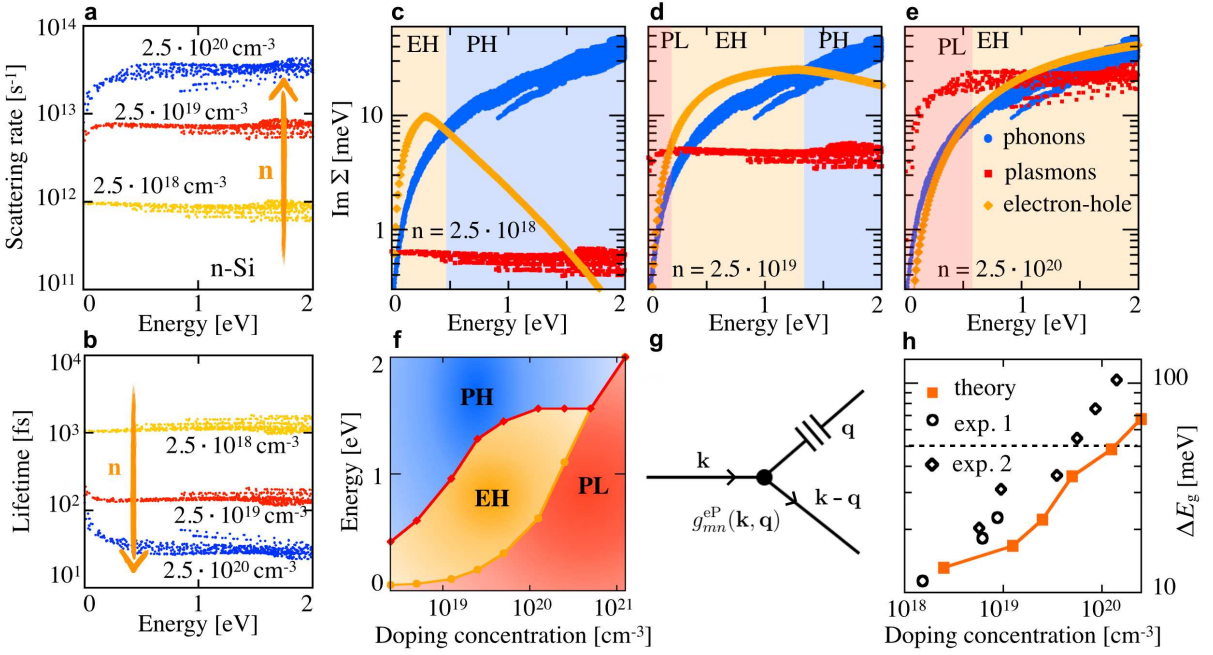}
  \caption{\label{fig2}
  (a) Calculated rates of electron scattering by plasmons, and (b) corresponding electron 
  lifetimes in doped silicon, for several carrier concentrations. The electron energy is referred to  
  the conduction band edge. (c--e) Comparison between the imaginary part of the
  electron-plasmon self-energy, the electron-phonon self-energy, and the self-energy associated
  with electron-hole pair generation. The carrier concentration increases from (c) to (e),
  and the electron energy is referred to the conduction band edge. 
  Shaded regions indicate the dominant scattering mechanism at a given electron energy, and
  `PL', `PH', `EH' stand for plasmons, phonons, and electron-hole pairs, respectively.
  (f) Energy vs.\ doping map of the largest contribution to the electron self-energy. 
  The energy is referred to the conduction band edge.
  (g) Diagrammatic representation of the electron plasmon scattering process. 
  (h) Calculated plasmon-induced band gap renormalization in silicon 
  as a function of carried density (orange squares and line), compared
  to the optical data from Ref.~\cite{aw/1991}  (experiment 1) 
  and Ref.~\cite{balkanski/1969} (experiment 2).
  The dashed horizontal line indicates the renormalization of the band gap by electron-phonon
  interactions, as reported by Ref.~\cite{zacharias/2015}.
  }
  \end{center}
  \end{figure*}

In order to investigate the effects of plasmons on the electronic structure we generalise
Pines' theory of \colora{electron-plasmon interactions in } the homogeneous electron gas \cite{pines1999elementary} to 
{\it ab initio} calculations for crystalline solids. 
\colora{ 
Our strategy consists of the following steps:
(i) We identify the energy vs.~wavevector dispersion relations of thermal plasmons.
This is achieved by determining the plasma energies from the 
poles of Im$\,\epsilon^{-1}({\bf q},\omega)$ for momenta below 
the critical wavevector $q_{\rm c}$ \cite{Nozieres1958}.
(ii) We single out the plasmonic contribution to the macroscopic dielectric function 
$\epsilon_{\rm M}$ via the Taylor expansion 
$\epsilon_{\rm P}({\bf q+G},\omega) = \left. \frac{\partial \epsilon_{\rm M} }
{\partial \omega} \right|_{\omega=\omega_{\rm P}({\bf q})} [\omega - \omega_{\rm P}({\bf q})]+i\eta$ 
in the vicinity of the plasmon frequency $\omega_{\rm P}({\bf q})$.
(iii) We calculate the electron-plasmon self-energy starting from 
many-body perturbation theory, and retain only the plasmonic screening.}
This leads to the retarded electron self-energy in Raleigh-Schr\"odinger perturbation theory \cite{Mahan2000}:
\begin{widetext}
  \begin{equation}\label{eq:sigma}
  \Sigma^{\rm eP}_{n{\bf k}} = \int\!\frac{d{\bf q}}{\Omega_{\rm BZ}}\sum_m 
     |g^{\rm eP}_{mn}({\bf k},{\bf q})|^2 
  \left[ \frac { n_{\bf q} + f_{m{\bf k+q}} } 
  {\varepsilon_{n{\bf k}} - \varepsilon_{m{\bf k+q}} + \hbar\omega_{\rm P}({\bf q}) + i\eta } 
  + \frac { n_{\bf q} + 1 - f_{m{\bf k+q}} } 
  {\varepsilon_{n{\bf k}} - \varepsilon_{m{\bf k+q}} - \hbar\omega_{\rm P}({\bf q}) + i\eta  } \right].
  \end{equation}
\end{widetext}
In this expression ${\bf k}$ and ${\bf q}$ are Bloch wavevectors, $m$ and $n$ band indices, 
$\varepsilon_{n{\bf k}}$
and $\varepsilon_{m{\bf k+q}}$ Kohn-Sham eigenvalues, $n_{\bf q}$ and $f_{m{\bf k+q}}$ Bose-Einstein
and Fermi-Dirac occupations, respectively, and $\eta$  a positive infinitesimal. The summation
runs over all states and the integral is over the Brillouin zone of volume $\Omega_{\rm BZ}$.
The quantities $g^{\rm eP}_{mn}({\bf k},{\bf q})$ represent the electron-plasmon scattering matrix elements
between the initial state $\psi_{n{\bf k}}$ and the final state $\psi_{m{\bf k+q}}$,
and are given by:
  \begin{equation}\label{eq:gs}
  g^{\rm eP}_{mn}({\bf k},{\bf q}) = 
  \left[\frac{\varepsilon_0\Omega}{e^2 \hbar} \frac{\partial\epsilon({\bf q},\omega)}
   {\partial\omega} \right]_{\omega_{\rm P}({\bf q})}^{-\frac{1}{2}}
   \frac{1}{|{\bf q}|}\langle \psi_{m{\bf k+q}} |e^{i{\bf q}\cdot{\bf r}} | \psi_{n{\bf k}} \rangle,
  \end{equation}
with $\Omega$ being the volume of one unit cell. Eqs.~(\ref{eq:sigma}) and (\ref{eq:gs}) are derived
in the Appendix. The present approach to electron-plasmon coupling in semiconductors is formally
identical to the standard theory of electron-phonon interactions \cite{Engelsberg1963}.
In particular, the $1/|{\bf q}|$ divergence of the electron-plasmon matrix elements at long wavelengths
is reminiscent of the Fr\"ohlich interaction between electrons and longitudinal-optical
phonons in polar semiconductors \cite{Froelich1954,Verdi2015}. This analogy is
consistent with the fact that bulk plasmons are longitudinal waves.
We now analyse the consequences of the self-energy in Eq.~(\ref{eq:sigma}).

From the imaginary part of the self-energy in Eq.~(\ref{eq:sigma})
we obtain the rate of electron scattering by thermal plasmons, using $\Gamma_{n{\bf k}}= 2\,{\rm Im}\, 
\Sigma_{n{\bf k}}/\hbar$.
Physically the two denominators in Eq.~(\ref{eq:sigma})
describe processes of one-plasmon absorption and emission, respectively. A 
diagrammatic representation of these processes is given in Fig.~\ref{fig2}g. Multi-plasmon processes
are not included in the present formalism, similarly to the case of electron-phonon 
interactions \cite{Engelsberg1963}, therefore we limit our discussion to low 
temperatures ($n_{\bf q}\ll 1$). 
Fig.~\ref{fig2}a shows the calculated electron-plasmon scattering rates in $n$-type silicon. The carrier
energies are referred to the conduction band edge. For standard doping levels ($n<10^{18}$~cm$^{-3}$)
the scattering rates fall below $10^{11}$~s$^{-1}$ as a result of the low intensity of the plasmon peaks
in Fig.~\ref{fig1}e, which is reflected in the strength of the matrix elements in Eq.~(\ref{eq:gs}).
However, at doping levels above 10$^{18}$~cm$^{-3}$, the strength of the
plasmon peak in the loss function increases considerably, and the frequency of scattering by thermal 
plasmons becomes comparable to electron-phonon scattering rates, 
10$^{12}$-10$^{14}$~s$^{-1}$ \cite{jalabert/1990,bernardi/2014}. Fig.~\ref{fig2}a
shows that at even higher doping levels these rates keep increasing by orders of magnitude,
and eventually dominate the cooling dynamics of excited carriers.

A complementary perspective on the carrier dynamics is provided by Fig.~\ref{fig2}b. Here we show
the electron lifetimes corresponding to the rates in Fig.~\ref{fig2}a, calculated as 
$\tau_{n{\bf k}} = 1/\Gamma_{n{\bf k}}$. Time-resolved reflectivity measurements of non-degenerate 
silicon ($n=10^{17}$~cm$^{-3}$ electrons photo-excited at 
$\sim$0.8~eV above the band edge) indicate thermalisation rates around 350~fs \cite{doany1988}.
In the same doping range our calculations yield plasmon-limited carrier lifetimes well above
10~ps, indicating that under these conditions electron-plasmon scattering is ineffective.
However, the scenario changes drastically for degenerate silicon, for which we calculate lifetimes
in the sub-picosecond regime. In particular, for doping levels in the range $10^{19}$-10$^{20}$~cm$^{-3}$
the electron-plasmon scattering reduces the carrier lifetimes to 25-150~fs. 
In these conditions electron-phonon and electron-plasmon scattering become competing mechanisms
for hot-carrier thermalisation.

In order to quantify the importance of electron-plasmon scattering we compare in Fig.~\ref{fig2}c-e the 
imaginary part of the electron self-energy associated with (i) electron-plasmon interactions, 
(ii) electron-phonon interactions, and (iii) and electron-hole pair generation. The methods 
of calculation of (ii) and (iii) are described in the Supplemental Materials \cite{sup}. From this comparison
we deduce that plasmons become increasingly important towards higher doping, and their effect
is most pronounced in the vicinity of the band edge. 
By identifying the largest contribution for each doping level and for each electron energy, we can construct the `scattering
phase diagram' shown in Fig.\ref{fig2}f. 
This diagram illustrates the regions in the energy vs.\ doping 
space where each scattering mechanism dominates. Unexpectedly in degenerate silicon
electron-plasmon scattering represents the dominant mechanism for hot-carrier relaxation.
This finding could provide new opportunities in the study of semiconductor-based plasmonics,
for example by engineering the doping concentration so as to selectively target the `plasmon region' 
in Fig.\ref{fig2}f.

We also evaluated the impact of electron-plasmon scattering processes on the carrier mobility
in silicon, by using the lifetimes computed above as a first approximation to the carrier 
relaxation times.
As shown in Fig.~S1 \cite{sup}, 
the explicit inclusion of electron-phonon scattering is essential to achieve 
a good agreement with experiment. On the other hand,  
were we to consider only electron-phonon scattering 
and electron-hole pair generation, we would overestimate
the experimental mobilities by more than an order of magnitude.

The real part of the electron self-energy in Eq.~(\ref{eq:sigma}) allows us to evaluate
the renormalization of the electron energy levels arising from the dressing of electron
quasiparticles by virtual plasmons. Since the renormalization of semiconductor band gaps 
induced by electron-phonon interactions attracted considerable interest lately 
\cite{Marini2008prl,Giustino2010prl,Marini2011prl,Botti2013prl,Monserrat2014prl,Antonius2015prl}, we here concentrate
on the quantum zero-point renormalization of the fundamental gap of silicon.
Computational details of the calculations and convergence
tests are {reported in the Supplemental Material \cite{sup}}.
Considering for definiteness a carrier density of $n=2.5 \cdot 10^{20}$~cm$^{-3}$, we find that
the electron-plasmon coupling lowers the conduction band edge by $\Delta E_c = -37$~meV at
zero temperature, 
and rises the valence band edge by $\Delta E_v = 30$~meV. 
\colora{ For carrier concentrations of $2.5\cdot10^{19}$~cm$^{-3}$ and $2.5\cdot10^{20}$~cm$^{-3}$ 
we verified that the BGN changes by less that 1~meV for temperatures up to 600~K (see Supplemental Material \cite{sup}).}
As a result at this doping concentration the band gap redshifts by $\Delta E_{\rm g} = \Delta E_c-\Delta E_v = -67$~meV.
This phenomenology is entirely analogous to the zero-point renormalization from electron-phonon
interactions \cite{Giustino2010prl}. Our finding is consistent with the fact that
the self-energy in Eq.~(\ref{eq:sigma}) and the matrix element in Eq.~(\ref{eq:gs}) are formally
identical to those that one encounters in the study of the Fr\"ohlich interaction. The doping-induced
band gap renormalization was also reported in a recent work on monolayer MoS$_2$ \cite{liang/2015},
therefore we expect this feature to hold general validity in doped semiconductors.
In order to perform a quantitative comparison with experiment, we show in Fig.~\ref{fig2}h our
calculated plasmonic band gap renormalization and measurements of the indirect absorption
onset in doped silicon \cite{balkanski/1969,aw/1991}. We can see that there is already good agreement 
between theory and experiment, even if we are considering only electron-plasmon couplings as the sole source
of band gap renormalization. Surprisingly the magnitude of the renormalization, 15-70~meV, is comparable
to the zero-point shift induced by electron-phonon interactions, 60-72~meV \cite{zacharias/2015}.

In summary, we presented an {\it ab initio} approach 
to electron-plasmon coupling in doped semiconductors.
We showed that electron-plasmon interactions are
strong and ubiquitous in a prototypical semiconductor such as doped silicon, as 
revealed by their effect on carrier dynamics, transport, and optical
properties.
This finding calls for a systematic investigation 
of electron-plasmon couplings in a wide class of materials.
More generally, a detailed understanding of the interaction between electrons and thermal plasmons 
via predictive atomic-scale calculations could provide a key into the design of plasmonic
semiconductors, for example by using phase diagrams such as in Fig.~\ref{fig2}f to tailor doping 
levels and excitation energies to selectively target strong-coupling regimes.
Finally, the striking similarity between electron-plasmon 
coupling and the Fr\"ohlich coupling in polar 
materials may open new avenues to probe 
 plasmon-induced photoemission 
kinks \cite{Damascelli2003rmp}, polaron satellites \cite{Moser2013prl,caruso/2015,caruso/2015/2},
as well as superconductivity, in analogy with the case of electron-phonon
interactions.\cite{Bustarret2006nature,Ekimov2004nature,Boeri2004prl,Pickett2004prl,Blase2004prl,Giustino2007prl}.

\acknowledgments 
F.C. acknowledges discussions with C. Verdi and S. Ponc\'e.
The research leading to these results has received funding from the Leverhulme Trust (Grant RL-2012-001),
the Graphene Flagship (EU FP7 grant no. 604391), the UK Engineering and Physical Sciences Research Council 
(Grant No. EP/J009857/1). Supercomputing time was provided by the University of Oxford Advanced Research 
Computing facility (http://dx.doi.org/10.5281/zenodo.22558) and the ARCHER UK National Supercomputing Service. 

\appendix 

\section{Electron self-energy for the electron-plasmon interaction}

Here we provide a derivation of the electron-plasmon coupling strength and the
self-energy [Eq.~(1) and (2)] by generalizing the theory of electron-plasmon interaction for the homogeneous electron gas
to the case of crystalline solids. 
We start from the electron self-energy in the $GW$ 
approximation \cite{Hedin1965,Aulbur/Jonsson/Wilkins:2000,Hybertsen1986}:
\begin{widetext}
  \begin{equation}\label{eq:sigmapw}
  \Sigma_{n\bf k} (\omega) = \frac{i \hbar}{2\pi} \sum_{m{\bf G}{\bf G'}}  \int\!\frac{d{\bf q}}{\Omega_{\rm BZ}}
  M^{mn}_{\bf G}({\bf k},{\bf q})^* M^{mn}_{\bf G'}({\bf k},{\bf q})
  \int \!d\omega' \frac{W_{{\bf G}{\bf G'}}({\bf q},\omega')}
  {\hbar\omega+\hbar\omega'+\mu -\tilde{\epsilon}_{m{\bf k+q}} },
  \end{equation}
where $M^{mn}_{\bf G}({\bf k},{\bf q})= \langle \psi_{m{\bf k+q}} |e^{i({\bf q+G })\cdot{\bf r}} |
\psi_{n{\bf k}} \rangle$ are the optical matrix elements, $\mu$ is the chemical potential,
and $\tilde{\epsilon}_{m{\bf k+q}}=\epsilon_{m{\bf k+q}} +i\eta\, { \rm sign}(\mu-\epsilon_{m{\bf k+q}})$.
The matrix $W_{{\bf G}{\bf G'}}({\bf q},\omega')= v({\bf q+G}) \epsilon_{{\bf G}{\bf G'}}^{-1}({\bf q}, \omega') $ 
represents the screened Coulomb interaction, and is
obtained from the bare Coulomb interaction $v({\bf q})=e^2/\varepsilon_0\Omega|{\bf q}|^2$  via the 
inverse  dielectric matrix $\epsilon_{{\bf G}{\bf G'}}^{-1}({\bf q}, \omega')$.
\colora{ The spectral representation of $W$ is given by:
\begin{align}\label{eq-chi-spec}
W_{{\bf G}{\bf G'}}({\bf q},\omega) 
& = \frac{ v({\bf q+G}) }{\pi} \int_0^\infty d\omega'
\frac{2\omega'}{\omega^2-(\omega')^2}{\rm Im\,}\epsilon_{{\bf G}{\bf G'}}^{-1}({\bf q}, \omega').
\end{align}
The dielectric matrix may be decomposed into:
\begin{align}
\epsilon^{-1}_{{\bf G}{\bf G'}}({\bf q},\omega) = 
\epsilon^{-1}_{\rm M}({\bf q+G},\omega) \delta_{\bf G G'} +
\epsilon^{-1}_{{\bf G G'}}({\bf q},\omega)(1 - \delta_{\bf G G'}).
\end{align}
where $\epsilon^{-1}_{\rm M}({\bf q+G},\omega)$ 
is the inverse macroscopic dielectric function. 
Since the plasmon energy $\hbar\omega_{\rm P}({\bf q})$ 
is defined by the condition $\epsilon_{\rm M}({\bf q+G},\omega_{\rm P}({\bf q})) = 0$, 
the plasmonic contribution to the dielectric matrix $\epsilon_{\rm P}$ 
can be singled out by Taylor-expanding $\epsilon_{\rm M}$ 
around the plasmon energy. 
Following Pines and Schrieffer \cite{pines1962} we have: 
\begin{align}\label{eq:eps-m-pl2}
\epsilon_{\rm P}({\bf q+G},\omega) &= 
\left. \frac{\partial \epsilon_{\rm M} }
{\partial \omega} \right|_{\omega=\omega_{\rm P}({\bf q})} [\omega - \omega_{\rm P}({\bf q})]+i\eta.
\end{align}
Making use of the identity $(a+i\eta)^{-1} = P(1/a) + i \pi \delta(a)$, 
and combining Eqs.~(\ref{eq:sigmapw}), (\ref{eq-chi-spec}), and (\ref{eq:eps-m-pl2})
yields the electron-plasmon self-energy:
\begin{align}\label{eq:sigmaplasmon-1}
\Sigma^{\rm eP}_{n\bf k}& (\omega) =
 \frac{i\hbar}{2\pi} \sum_{m{\bf G}}  \int\!\frac{d{\bf q}}{\Omega_{\rm BZ}} 
 |M^{nm}_{\bf G}({\bf k},{\bf q})|^2 
\int d\omega'
\frac{2\omega_{\rm P}({\bf q})}{\omega'^2-[\omega_{\rm P}({\bf q})]^2}
\left[\left.\frac{\partial\epsilon_{\rm M}}{\partial\omega}
\right|_{\omega=\omega_{\rm P}({\bf q})}\right]^{-1} 
\frac{ v({\bf q+G})}
{\omega+\omega'+\mu -\tilde{\epsilon}_{m{\bf k+q}} }.
\end{align}
This expression may be recast into the form of a self-energy describing the 
interaction between electrons and bosons in the Migdal approximation}
\cite{Engelsberg1966,Lundqvist1967,Langreth1970,Overhauser1971,Mahan2000}: 
  \begin{equation}\label{eq:sigma-pl-final}
  \Sigma^{\rm eP}_{n\bf k} (\omega) =
  \frac{i\hbar}{2\pi} {\sum}_m \int \!\frac{d{\bf q}}{\Omega_{\rm BZ}} \int \! d\omega'\,
  |g^{\rm eP}_{mn}({\bf k},{\bf q})|^2 
   D_{\bf q}(\omega')\, G_{m{\bf k+q}}(\omega+\omega').
  \end{equation}
\end{widetext}
\colora{
Since for doped semiconductors $q_{\rm c}$ is typically within the first Brillouin zone,
we dropped the dependence on the reciprocal lattice vectors ${\bf G}$.}
The matrix elements appearing in this expression are defined in Eq.~(2); 
$G$ represents the standard non-interacting (Kohn-Sham) electron Green's function, $G_{n{\bf k}}(\omega) =
[\hbar\omega-(\tilde{\varepsilon}_{n{\bf k}}-\mu)]^{-1}$, and we introduced the
`plasmon propagator': 
$D_{\bf q}(\omega) = 2\omega_{\rm P}({\bf q})/[\hbar(\omega^2-\omega_{\rm P}^2({\bf q}))]$.
Equation~(\ref{eq:sigma-pl-final}) represents the prototypical electron self-energy arising
from electron-boson interactions. From this expression the result in Eq.~(1)
follows by standard integration in the complex plane \cite{Mahan2000}.

For completeness we note that Eq.~(\ref{eq:sigma-pl-final}) can also be derived 
from the electron-boson coupling Hamiltonian $\hat {H}^{\rm eP} =\Omega_{\rm BZ}^{-2}
\sum_{n m}\int d{\bf k} d{\bf q} \, g^{\rm eP}_{nm}({\bf k},{\bf q})
\hat c_{m{\bf k+q}}^\dagger \hat c_{n{\bf k}} (\hat b_{\bf q} + \hat b^{\dagger}_{\bf -q})$,
where $\hat b^{\dagger}_{\bf -q}$ ($\hat b_{\bf q}$) and $\hat c_{m{\bf k+q}}^\dagger$ ($\hat c_{n{\bf k}}$)
are the boson and fermion creation (destruction) operators, respectively.
\colora{
As a consistency check, we note that the electron-plasmon coupling coefficients Eq.~(2)
reduce to the results of Pines and Schrieffer for homogeneous systems \cite{pines1962}. 
In particular, for an homogeneous electron gas we have 
$M^{nm}_{{\bf G}}({\bf k},{\bf q})=\delta_{nm}$ and 
$\epsilon_{\rm M}= 1 - {\omega_{\rm P}({\bf q})/ \omega^2 }$. 
In this case, the partial derivative in the definition of the electron-plasmon coupling 
coefficients can be evaluated analytically, giving 
the results of Ref.~\cite{pines1962}, 
$g^{\rm eP}({\bf q}) = ({2\pi e^2 \hbar \omega_{\rm P}({\bf q})}/{\epsilon_0 q^2})^{\frac{1}{2}}$.

Finally, we emphasize that the structure of Eq.~(\ref{eq:sigmaplasmon-1}) 
stems directly from the identification of the plasmonic contribution
to the dielectric function through the linearization of Eq.~(\ref{eq:eps-m-pl2}), and it is 
reflected in the inclusion of the plasmon oscillator strength
$\left.\frac{\partial\epsilon_{\rm M}}{\partial\omega}
\right|_{\omega=\omega_{\rm P}({\bf q})}$ in the coupling coefficients [Eq.~(2)].
This procedure distinguishes the electron-plasmon self-energy 
from the conventional $GW$ self-energy in the plasmon-pole approximation, 
and justifies its application to the study of thermal plasmons in 
doped semiconductors. 

\section{Plasmon damping}

To investigate the effects of extrinsic carriers on thermal plasmons, 
we consider the Fermi golden rule for the rate of change of the plasmon 
distribution function \cite{pines1962}:
\begin{widetext}
\begin{align}\label{eq-plasmondecay}
R_{\bf q} &= \frac{2\pi}{\hbar} 
\sum_{\bf k}^{\rm BZ} 
\int\frac{d{\bf k}}{\Omega_{\rm BZ}}
\sum_{nm} 
|g^{\rm eP}_{mn}({\bf k},{\bf q})|^2 
[ ( n_{\bf q} + 1 ) f_{n{\bf k+q}} ( 1- f_{m{\bf k}})   
-  n_{\bf q} f_{m{\bf k}} (1- f_{n{\bf k+q}} )
] \delta(\epsilon_{m{\bf k}} +\hbar\omega_{\rm P}({\bf q}) - \epsilon_{n{\bf k+q}})   
\end{align}
\end{widetext}
where $\hbar\omega_{\rm P}({\bf q})$ are plasmon energies, 
$g^{\rm eP}$ electron-plasmon coupling coefficients,  and $n$/$f$ are 
Bose/Fermi occupation factors for plasmons/electrons. 
In practice, the first term accounts for the increase of the 
plasmon population induced by the absorption of an electron-hole pair, 
whereas the  inverse process is described by the second term.
Thermal plasmons are well defined for momenta smaller that 
the critical momentum cutoff given by the wavevector: 
$q_{\rm c} = k_{\rm F}\left[(1+\hbar\omega_{\rm P}/\varepsilon_{\rm F})^{1/2}-1\right]$, 
with $k_{\rm F}$ and $\varepsilon_{\rm F}$ being
the Fermi wavevector and the Fermi energy, respectively.
By definition (see, e.g., \cite{pines1999elementary}) $q_{\rm c}$ is the smallest 
momentum satisfying the condition 
$\hbar\omega_{\rm P}({\bf q}) = \epsilon_{n{\bf k+q}} - \epsilon_{m{\bf k}}$. 
Thus for $q<q_{\rm c}$, the Dirac $\delta$ in Eq.~(\ref{eq-plasmondecay}) vanishes, indicating that, 
while excited carriers may decay upon plasmon emission, the inverse processes, 
whereby a thermal plasmon decays upon emission of an electron-hole pair, 
is forbidden. Therefore, thermal plasmons are {\it undamped} by 
other electronic processes, and their decay for $q<q_{\rm c}$ may be 
ascribed exclusively to plasmon-phonon and plasmon-plasmon scattering. 

  \begin{figure}[t]
  \begin{center}
  \includegraphics[width=0.4\textwidth]{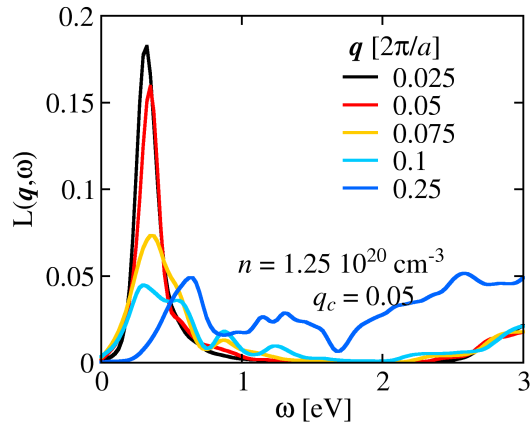}
  \caption{\label{fig:landau} Momentum dependence of plasmon peak in the loss function of silicon at 
  a doping concentration of $1.25\cdot10^{20}$~cm$^{-3}$, corresponding to 
  a critical momentum cutoff $q_{\rm c}=0.05$ in $2\pi/a$ units. }
  \end{center}
  \end{figure}

  \begin{figure*}[t]
  \begin{center}
  \includegraphics[width=0.75\textwidth]{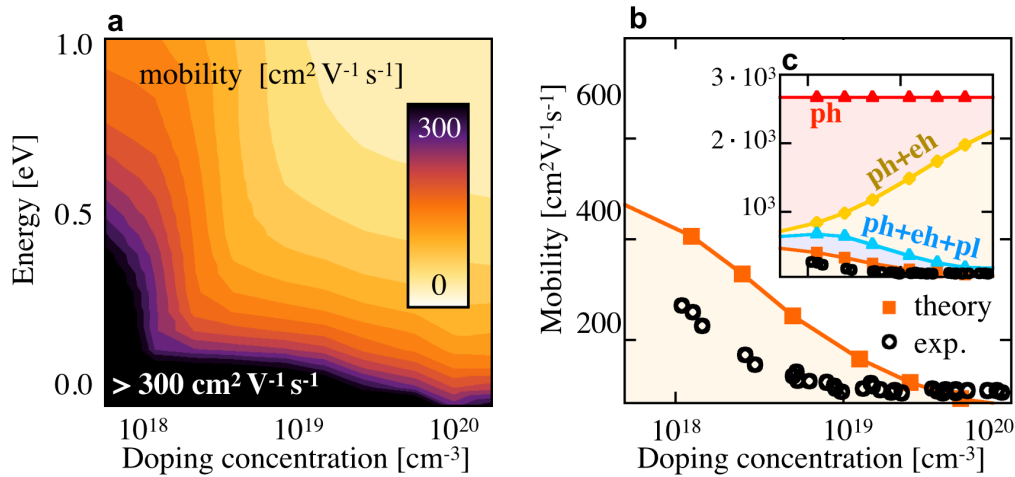}
  \caption{\label{fig3}
  (a) Calculated electron mobility in $n$-type silicon, as a function of carrier density and
  energy relative to the chemical potential. (b) Comparison between calculated and measured
  electron mobilities in silicon as a function of doping. The black circles indicate experimental
  low-temperature mobility data from Ref.~\cite{mousty/1974}. The orange squares and line
  represent our complete calculation including electron-plasmon (pl), electron-phonon (ph),
  electron-hole (eh), and impurity scattering. (c)  Partial contributions to the mobility are
  shown as red (ph), and yellow (ph+eh).}
  \end{center}
  \end{figure*}

To exemplify the effect of Landau damping on the plasmon dispersion, 
we illustrate in Fig.~\ref{fig:landau} the plasmon peak 
in the loss function of silicon at a doping concentration of 
$1.25\cdot10^{20}$~cm$^{-3}$. At these carrier concentration, we obtain a 
momentum cutoff $q_{\rm c}=0.05$ in units of $2\pi/a$, with $a$ being the 
lattice constant. For $q<q_{\rm c}$, the loss function exhibit well 
defined plasmon peak with a peak intensity larger than the continuum of 
electron-hole excitations. 
For $q>q_{\rm c}$, on the other hand, the plasmon intensity is reduced as a 
consequence of the lifetime effects introduced by Landau damping, and its 
intensity becomes essentially indistinguishable from the spectral signatures of 
electron-hole pairs. 
}

\section{Plasmon-limited mobility}

We now evaluate the impact of electron-plasmon scattering processes on the carrier mobility
in silicon. In the relaxation-time approximation the mobility is given by
$\mu= e\tau^{\rm tot} /m_{\rm e} m^*$, where $m^*$ is the conductivity effective mass,
that is the harmonic average of the longitudinal and transverse masses, 
and $\tau^{\rm tot}$ is the  scattering 
time arising from processes involving plasmons (eP), phonons (ep), 
electron-hole pairs (eh), and impurities (i).
Noting that scattering time and relaxation time
differ by less that 10\% at low carrier concentrations \cite{dassarma/1985},
we follow Matthiessen's rule to calculate $\tau^{\rm tot}_{n{\bf k}} = \hbar / 2 \,
{\rm Im}(\Sigma_{n{\bf k}}^{\rm ep} + \Sigma_{n{\bf k}}^{\rm eP} 
  + \Sigma_{n{\bf k}}^{\rm eh} + \Sigma_{n{\bf k}}^{\rm i})$,
where $\Sigma_{n{\bf k}}^{\rm ep}$, $\Sigma_{n{\bf k}}^{\rm eh}$, and $\Sigma_{n{\bf k}}^{\rm i}$  
are the electron self-energies associated with each interaction. 

Strictly speaking the mobility $\mu$ 
is an average property of all the carriers in a semiconductor; however, for illustration purposes,
it is useful to consider a `single-electron' mobility obtained as $\mu_{n{\bf k}}= e\tau^{\rm tot}_{n{\bf k}} 
/m_{\rm e} m^*$. This quantity is shown in Fig.~\ref{fig3}a. In this figure we see that
the mobility decreases as one moves higher up in the conduction band; this behavior 
relates to the increased phase-space availability for electronic transitions. 
In addition we see that the mobility decreases with increasing carrier concentration.
In order to analyse this trend we give a breakdown of the various sources of scattering
in Fig.~\ref{fig3}c, and we compare our calculations to experiment.
Here we show the carrier mobility at 300~K averaged on the Fermi surface defined by the
doping level. 
Electrical measurements at high doping \cite{mousty/1974} 
yield mobilities in the range of 100-300~cm$^2\,$V$^{-1}$s$^{-1}$
for carrier densities between $10^{18}$ and $10^{20}$~cm$^{-3}$;
these data are shown as black circles in Fig.~\ref{fig3}b-c. Were we to consider only
electron-phonon scattering and electron-hole pair generation, we would overestimate
the experimental mobilities by more than an order of magnitude (red and yellow lines
in Fig.~\ref{fig3}c). Impurity scattering reduces this discrepancy to some extent, 
but there remains a residual difference at the highest doping levels.
It is only upon accounting for electron-plasmon scattering that the calculations
exhibit a trend in qualitative and even semi-quantitative agreement with experiments
throughout the entire doping range. In particular the scattering by plasmons is key
to explain the anomalous low mobility of 100~cm$^2\,$V$^{-1}$s$^{-1}$ above $n=10^{19}$~cm$^{-3}$.
Even through the inclusion of electron-plasmon scattering a residual discrepancy
between theory and experiment is still observed, which we ascribe to the simplified
models adopted in the description of electronic scattering with electron-hole pairs and
impurities.
This observation leads us to suggest that the origin of the 
mobility overestimation in earlier calculations could be connected
with the neglect of electron-plasmon scattering \cite{meyer1987,restrepo2009}.

\end{document}